\documentclass[12pt,preprint]{aastex}

\slugcomment{to be submitted to the Astrophysical Journal Letters.}

\shorttitle{The source of the TeV emission of M87}
\shortauthors{Georganopoulos, Perlman, Kazanas}

\begin{document}

\title{ Is the core of M87 the source of its TeV emission?  Implications for unified schemes.}

\author{Markos Georganopoulos\altaffilmark{1,2}
Eric S. Perlman \altaffilmark{1}, 
Demosthenes Kazanas\altaffilmark{2}}

\altaffiltext{1}{Department of Physics, Joint Center for Astrophysics, University of Maryland-Baltimore County, 1000 Hilltop Circle, Baltimore, MD 21250, USA}
\altaffiltext{2}{Laboratory for High Energy Astrophysics, 
NASA Goddard Space Flight Center, 
Code 661, Greenbelt, MD 20771, USA.}

\begin{abstract}

M87 has been recently shown to be a TeV source which is likely to be variable.
Based on this, and on contemporaneous optical and X-ray monitoring, 
we argue that the source
of the TeV emission is the  core of M87 and not one of two jet knots
(HST-1 and A) with X-ray brightness  comparable to that of the core. 
We model the TeV emission in the core as inverse Compton (IC) emission from the
 base  of the jet.
Homogeneous models  fail to  reproduce the spectral
energy distribution (SED) and, in particular, the  TeV flux. 
They also fail to  comply with the unified scheme 
 of  BL Lacertae (BL) objects and FR I  radio galaxies. 
A jet that decelerates from a Lorentz factor  $\Gamma\sim 20$
down to $\Gamma\sim 5$ over a length of $\sim 0.1  $ pc reproduces the 
observed SED  of the M87  core, and, when aligned to the line of sight, 
produces a SED similar to those of TeV BLs.
The TeV flux in the decelerating jet model is successfully reproduced as 
  upstream Compton (UC)  scattering, a recently identified emission mechanism, 
in which energetic  electrons of the upstream faster flow upscatter the low 
energy photons  produced in the slower downstream part of the flow.

\end{abstract}

\keywords{ galaxies: active --- quasars: general --- quasars: individual (M87) --- radiation mechanisms: 
nonthermal --- X-rays: galaxies}

\section{Introduction \label{intro}}

% BACKGROUND INFO ON GAMMA RAY EMISSION AND UNIFICATION

Gamma-ray emission in  GeV energies  has been detected in
several  bright BLs  \citep{hartman99}.
A handful  of nearby, bright BLs have been also detected in the
TeV regime (e.g. Aharonian et al. 2005).
%, and their number is expected
%to increase with the advent of the next generation of TeV telescopes 
%({\sl HESS, MAGIC, VERITAS}).
  Gamma-ray emission at some level is 
expected   from radio galaxies, since, according to the unified 
scheme (e.g. Urry \& Padovani 1995), BLs are  FR I  radio galaxies 
\citep{fanaroff74}  with their relativistic jets oriented toward the observer. 
So far, the only radio galaxies  detected  in Gamma-rays are the FR I's 
Cen A \citep{sreekumar99} and M87  ({\sl HEGRA}, Aharonian et al.  2003; 
{\sl HESS},  Beilicke et al. 2005) at TeV energies.
 Also, a   tentative GeV detection of  NGC 6251 \citep{mukherjee02} 
has been strengthened by its  hard X-ray detection. 
(Foschini et al 2005).
% SPECIFICS OF M87 AT RADIO-XRAY FREQUENCIES, INCLUDING KNOT HST-1.

M87 is among the nearest galaxies with a bright 
radio/optical/X-ray jet.  Due to its proximity (distance 16 Mpc)
the jet has been  studied with unparalleled spatial
 resolution  (e.g., Perlman et al. 1999, 2001 and references therein).  
The synchrotron emission extends to X-ray energies in all 
jet knots, as well as the core 
(e.g., Perlman et al. 2001,  Marshall et al. 2002, Wilson \& Yang 2002, Perlman \& Wilson 2005). 
{\sl HST} observations of the core   \citep{tsvetanov98} showed 
variability on  timescales of  $\sim 1$ month.
{\sl HST} and {\sl Chandra} semi-monthly monitoring of  the jet  during 
the last three years (Harris et al. 2003, 2005; Perlman et al. 2003)
 established the broadband nature of  this variability.
Surprisingly, these observations  detected an enormous 
(factor $\sim 50$) ongoing flare in the jet component HST-1, $0.8''$ 
(60 pc projected)  from the core, with similar variability timescales.  
%At the time of writing HST-1 is  more luminous than the core in both 
%the optical/UV and X-rays by factors $\gtrsim 2-5$, depending on the band. 

% POSING THE PROBLEM

The discovery of flaring optical and X-ray emission
  established an important link between M87 and the 
BLs, as did the detection of apparent superluminal motion in the jet 
at radio \citep{biretta95}  and  optical \citep{biretta99} frequencies.
The apparent superluminal velocities in the optical 
of $u_{app}\approx 6 $ c require a jet  orientation angle 
$\theta \lesssim 19^{\circ}$   and $\Gamma\gtrsim 6$. 
The detection of  TeV  emission 
constitutes another link between FR I's  and BLs.
The TeV  data appear to indicate  significant variability, 
as the 1998/99 ({\sl HEGRA}) and 2003/04 ({\sl HESS})
fluxes differ  by over a factor 3  at 
$>3 \sigma$ significance (Beilicke et al. 2005).
In BLs,  correlated X-ray and TeV variations
(e.g. Takahashi et al. 1996)  suggest that the TeV emission is produced
at the core.  
In the case of M87, it is necessary to consider as possible  TeV emitters
three  components with comparable optical and X-ray synchrotron luminosity:
the core, HST-1, and knot A, an extended jet feture located $12.4''$ from the core.

% THE SCOOP

Based on the likely TeV variability, we argue in \S 2
 that the most plausible source of the TeV emission
is the core and  not  HST-1 or knot A.
 In \S 3 we consider the modeling constraints imposed by the 
core SED  and the unified scheme and
show that homogeneous models cannot satisfy these constraints. 
We then show  (\S 3.1) that  a decelerating relativistic jet can solve 
these problems.
Finally, in \S 4 we sum up our conclusions.

\section{The source of the TeV emission: Variability constraints.}

% WE REQUIRE: CORRELATED VARIABILITY

The production of TeV photons requires  TeV energy electrons, 
which also  emit  optical -- X-ray synchrotron  radiation,
resulting in correlated TeV and optical -- X-ray variability.  
The possible TeV emission sites are,  therefore, 
the brightest optical -- X-ray features of  the M87 jet. 
These are the core and knots 
A and HST-1, all of comparable optical and X-ray fluxes 
(within a factor $\sim 5-10$). 
We now use the fact that  the TeV emission has likely dropped 
by a factor of $\sim 3$ between 1998/99 and 
2003/04, to constrain the actual TeV source.

% IT IS NOT KNOT A

We first consider knot A, one of  the largest bright regions in the M87 jet, 
with its  flux maximum  region  $\approx 1''
\times 0.7''$  in {\sl HST} \citep{sparks96} and {\sl Chandra}
\citep{perlman05}  images, translating to 80 $\times$ 55
pc (projected).  
Such a large region cannot 
vary on timescales of a few years, and, indeed, 
monitoring  by both
ROSAT \citep{harris99} and {\sl Chandra} \citep{harris05} does not 
detect variability of more
than $\sim 10-20\%$ in the X-rays.   

% IT IS NOT HST-1

We turn now to knot HST-1, which is currently
exceeding the  core flux in the optical, UV, and X-rays by factors
$\gtrsim 2-5$, depending on the band (Harris et al. 2003, 2005; Perlman et al. 2003). 
%
%Given that its X-ray emission is synchrotron radiation,  some level
%of TeV emission  is expected.
%
In Spring
2004, the X-ray flux of HST-1  was $\sim 3$ times brighter than it was in 
Spring 2003, and since 1998/99 the
increase is far larger --  nearly a factor 50.  By comparison, there is no
evidence of  an increase in TeV flux from 2003 to 2004, or  from
1998 to 2004 \citep{aharonian03,beilicke05}. 
Since the X-ray and 
TeV variations are not  correlated, the contribution of HST-1 to the TeV 
output of M87 seems not to be significant. 
This may indicate a low synchrotron photon 
compactness for HST-1, which can be used to set, in a manner similar to 
that of Stawartz et al. (2004) for knot A, a 
lower limit on  its  magnetic field.
% IT IS THE CORE 

 Having excluded knots A and HST-1, we turn to the core. {\sl  HST, ROSAT} 
and {\sl Chandra} monitoring (Harris et al. 1999, 2003, priv. comm.; Perlman et al. 2003)
 suggest that its  variations may be similar  to
those observed by HEGRA and HESS in 1998/99, and 2003/04.
We therefore believe that the most likely source for the  TeV emission 
 is the  core of M87.

\section{ Modeling  the   core SED}

%\subsection{Spectral and Unification Constraints on  Homogeneous  Models}

% APRIL 2003 SED

\begin{figure}
\epsscale{0.8}
\plotone{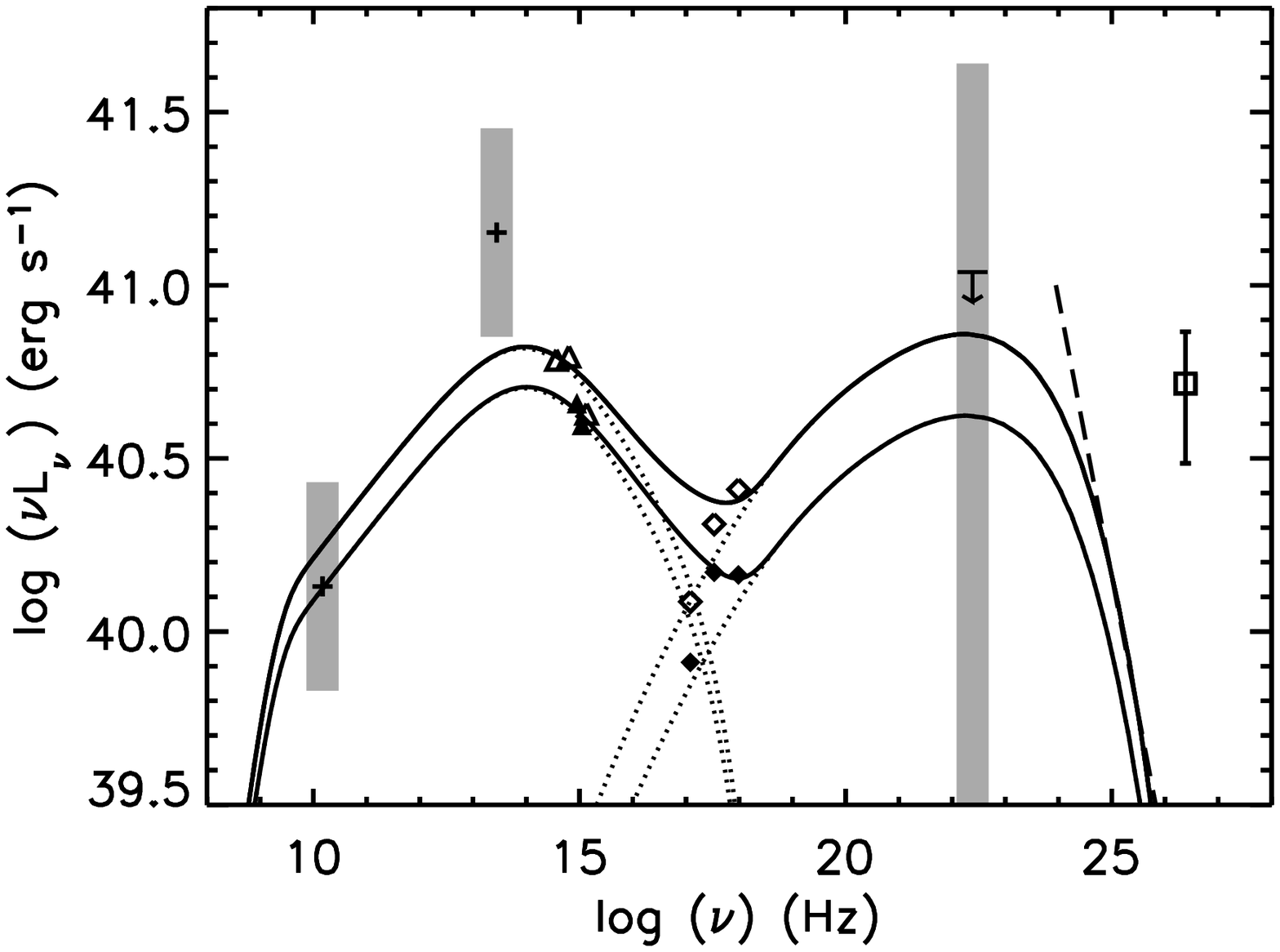}
\caption{Open symbols: the core SED in April 2003;  optical \citep{perlman03},
  X-ray \citep{harris03, perlman03} 
and TeV \citep{beilicke05}.  Solid  symbols: early May  2003 SED;
 \citep{perlman03, harris03}.  Crosses: non-contemporaneous radio
 \citep{biretta91} and 10 $\mu m$  \citep{perlman01} data (the gray 
bands indicating flux  variations  by a factor of 2). 
We also plot the $2\sigma$ {\sl EGRET} upper
limit 	\citep{sreekumar96}, with the gray band reaching up to
a flux 4 times larger than the detection limit. The dotted 
(synchrotron and SSC) and solid (total) lines 
correspond to the model SEDs of a homogeneous model  
resulting for $R=10^{17} cm$, $B=0.03$ G, $\delta=1.4$, 
derived from the SED using the formalism of  \S 3.
The two SEDs differing only  in the jet comoving power 
$L_{jet}$, with    $L_{jet}=7.5 \times 10^{42} $ erg $s^{-1}$ 
for the lower one and  $L_{jet}= 10^{43} $ erg $s^{-1}$ for the higher one. 
The broken line indicates the analytical slope of the 
SSC spectrum in the KN regime.}
\label{m87sed}
\end{figure}

Given the  $\sim 1$ month  variability timescale of the core emission, 
its modeling requires   contemporaneous broadband observations.
The most complete coverage of the core SED is that of April and May 2003 
%with  optical \citep{perlman03},  X-ray \citep{harris03, perlman03} 
%and TeV \citep{beilicke05} fluxes 
(Figure \ref{m87sed}).
We include  non-contemporaneous radio and IR data
(crosses), given that the amplitude of the  core variability does not 
 exceed a factor of $\sim 2$ amplitude.
 We also plot the $2\sigma$ {\sl EGRET} upper
limit 	\citep{sreekumar96}, with the gray band reaching up to
a flux 4 times larger than the detection limit, as expected when  
the IC seed photons are the source's 
synchrotron photons:
 a change  in the electron injection  results respectively in linear and 
 quadratic changes of the synchrotron and IC fluxes.

% THE SYNCHROTRON SED PEAK CORRESPONDS TO A COOLING BREAK 
 
The synchrotron  peak between  $10^{12}-10^{14}$ Hz  corresponds to
 a  cooling break rather that to a cutoff of the electron energy
 distribution  (EED), because this component exhibits a power-law shape that 
extends
 to X-ray energies (e.g. Marshall et al. 2002). This can be  seen in the
    2003 spectral data \citep{perlman03, harris03}.  The  decrease in the 
synchrotron  SED from April to May 2003 must  be accompanied by a quadratic
 decrease in the IC SED.   If there was no synchrotron
contribution in the X-rays, the X-ray spectral slope should have remained
 constant, with only a reduction of the X-ray flux.
However, the X-ray slope steepens, suggesting  a significant 
synchrotron contribution in the low state. 
%The fact that the synchrotron SED peak corresponds to a cooling break is a  very significant modeling constraint. 
Note that an extrapolation of the two higher
energy fluxes (1.4 and 4.0 KeV) down to 0.5 KeV  overproduces the 0.5 KeV
flux in  both April and May 2003. 
This may be due to  an intrinsic 
absorption component \citep{perlman05}.

The SED of the core of M87, together  with the  FR I - BL unification 
(Urry \& Padovani 1995) argue against a
homogeneous model for the core.  The synchrotron 
SED of TeV BLs is
%very different from those of the M87 core. The common 
%characteristics of TeV BL spectra, after taking into account the absorption 
%of the TeV emission from the diffuse infrared background, include: (a) An IC
%peak in the SED at 1-10 TeV. (b) A Compton dominance  $C_d=L_C/L_S$
%(ratio of IC to synchrotron  luminosity)  ranging between  $C_d \lesssim  1$ 
%for MKN 421 (e.g. Konopelko et al. 2003),  up to  $C_d \gtrsim 10 $ for the 
%most distant (redshift $z=0.129$)   TeV BL 1ES 1426+428 
%(Costamante et al. 2003). (c) 
characterized by a spectral break at $\sim 10^{15-16} $ Hz, 
with  the  SED  still increasing  above this break, 
to  reach  its peak at $\sim 10^{17}-10^{19}$ Hz, 
beyond which it exhibits a cutoff (e.g. the 1997 flare of MKN 501, Krawczynski 
et al. 2000). This is in strong contrast with the declining IR -- optical -- 
X-ray SED of the core of M87. 
Because relativistic motion preserves the SED
shape, simply shifting it along the frequency and luminosity axes, a
 simple boosting
 cannot shift the SED of the core of 
M87 into  the mold of the SED of the TeV BLs.

Even if M87 is a special case that,  if aligned,
 does not  resemble the TeV BL Lacs, we still need to reproduce
 its SED. This turns out to be practically impossible for the homogeneous 
model. To sketch why this is the case, we will make use of four  quantities
derived or constrained by observations:
the synchrotron peak frequency $\nu_s=10^{12-14}$ Hz,
 the synchrotron peak luminosity $L_p\approx 10^{41}$ erg s$^{-1}$ ,
 the Compton dominance (ratio of IC to synchrotron  
luminosity) $C_d\sim 1- \mbox{few}$, and the variability timescale
 $t_{var} \sim 1$ month.  The bolometric synchrotron luminosity 
is  $L_{s,bol}=L_p k_b$, where  $k_b=2/[(3-p)(p-2)]$ and 
$p$ is the  index of the injected power law electron distribution.
If $\gamma_b$ is the Lorentz factor  where the cooling time equals the
escape time, $\gamma_b\propto B^{-2} R^{-1}$, where $B$ is the magnetic field
and $R$ is the radius of the source, and $\nu_s\propto \delta B\gamma_b^2 
\propto \delta B^{-3} R^{-2}$,  where $\delta$ is the usual Doppler factor.
Noting that $C_d=U_s/U_B$, 
where $U_s\propto L_{s,bol} R^{-2}\delta^{-4}$ is the 
synchrotron photon energy density
and $U_B=B^2/(8\pi)$ is the magnetic field energy density, we have 
$C_d\propto L_p k_b R^{-2}B^{-2}\delta^{-4}$. 
Using these expressions for $\nu_s$ 
and $C_d$, together with $R\sim c t_{var} \delta $, we obtain
$B\propto \delta^5 C_d \nu_s^{-1} L_p^{-1} k_b^{-1}$
and 
$\delta^8 \propto \nu_s  L_p^{3/2}k_b^{3/2}C_d^{-3/2}t_{var}^{-1}$.
 The last expression permits  only   
a very narrow range  $\delta\sim 1-2$, even after allowing for a generous
range in the SED-derived parameters.
%, including    a variation of  $ \nu_s =10^{12- 14}$ HZ (the most uncertain SED-derived parameter). 
Using the above  and  noting that the peak of the
 synchrotron self Compton (SSC) component is at a frequency
$\nu_{SSC}\approx \nu_s \gamma_b^2$, we obtain
\begin{equation}
\nu_{SSC}\approx 1.4 \times 10^{23} \delta^2 \nu_{s,14}^2 t_{var, m} \left ({ C_d \over L_{p,41}k_b}\right)^{1/2} \;\;\mbox{Hz},
\end{equation}
where $\nu_{s,14}= \nu_s / 10^{14}$,  $L_{p,41}=L_p/ 10^{41}$, and $t_{var,m}$ is
the variability timescale in months.
For a  SED -- derived set of parameters ($ L_{p,41}=1$, $ C_d=1$, $ k_b=8.33$  for $p=2.6$ required by  spectral fitting), 
the SSC component cannot peak at  
$\nu_{SSC} \gtrsim 10^{23} $ Hz, even for  
$\nu_s=10^{14}$ Hz.

While at $\nu_{SSC}$  the emission is in the Thomson regime,
due to  the reduced Klein-Nishina cross-section the SSC luminosity
$L_{SSC,KN}$ at higher 
frequencies  drops  much faster, $L_{SSC,KN} \propto \nu^{-(p-1)/2}$
(Georganopoulos et al., in prep),  
than anticipated in the Thomson case, $L_{SSC,T} \propto \nu^{-(p-2)/2}$ .
Because of this steeper SED at the upper end of the SSC component and
the  upper limit on $\nu_{SSC}$ (see Eq. (1) above), homogeneous models 
fail  to match the observed TeV flux.
A typical example  of this is shown in Figure \ref{m87sed}, where we plot the
SED for two different jet powers. 
%Note in both cases the steep slope of the
%model TeV spectrum is  in good agreement with the analytical slope.
Note  how in the low state the  stronger reduction of the SSC SED  
results in a steeper X-ray spectrum, in agreement with the observations.

\subsection{Decelerating flow and Upstream Compton scattering}

 It has been noted (Georganopoulos \& Kazanas 2003b, GK03b) that homogeneous  
models can fit the SED of TeV BLs only by 
invoking very high Doppler factors 
($\delta\sim 50$; e.g. Krawczynski, Coppi, \& Aharonian 2002).
Similarly fast flows (Perlman et al. 2003) are required by homogeneous models
 if one wants to relate 
the M87 variability timescale ($\sim  1 $ month) with those of TeV blazars 
($\sim$ few hours). 
These values are in strong disagreement with
the BL -- FR I unification, 
because they severely underpredict the observed
FR I luminosities  and overpredict their number counts
 \citep{chiaberge00,trussoni03}. They  also  conflict
with the  slow  apparent motions ($u_{app}\lesssim  c$) 
in the pc-scale  jets of TeV BLs 
\citep{marscher99, piner04, piner05}. This motivated GK03b to propose that 
  TeV BL jets  decelerate at sub-pc scales from an  initial Lorentz factor 
$\Gamma_0\sim 10-20$ down to $\Gamma\sim$ few.
In such flows, electrons are more energetic in 
the faster base  of the flow and are radiatively cooled to lower energies 
as they  advect   downstream where the jet decelerates.
Beaming, hence,  is frequency dependent. 
%and 
%cooling spectral breaks different than the standard 
%value of $\Delta\alpha=1/2$  are naturally produced, with  $\Delta\alpha$ 
%increasing with increasing jet orientation angle $\theta$. 
In addition to this, 
upstream Compton (UC) scattering, a newly identified
mechanism \citep{georganopoulos03a} in which  the energetic electrons of 
the upstream fast flow  
'see' the  seed photon field produced by the lower energy electrons of 
the slower  downstream flow boosted, increases the level and beaming 
of high energy Compton emission.

\begin{figure}
\epsscale{0.8}
\plotone{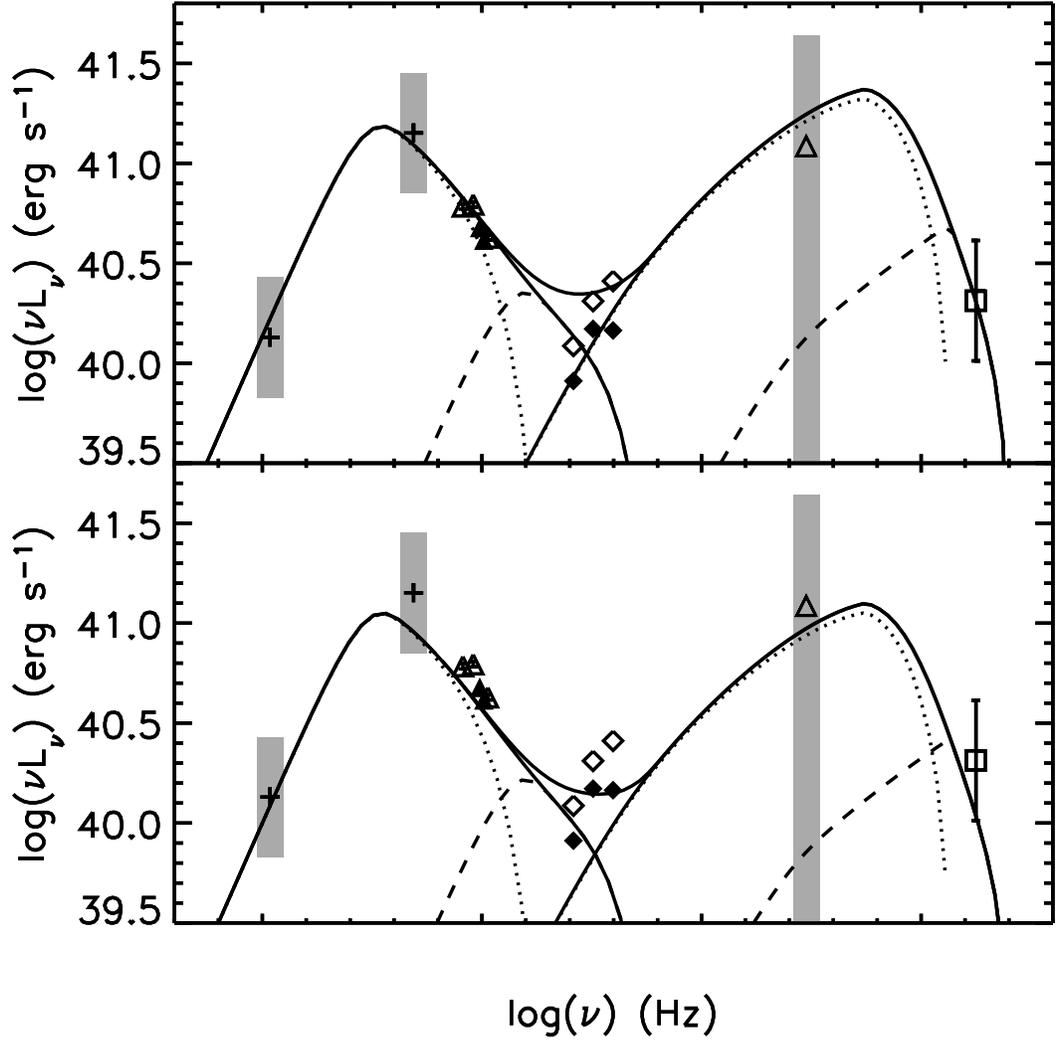}
\caption{Upper panel: The SED of a decelerating flow seen under an 
angle $\theta=13^{\circ}$.  The flow decelerated from $\Gamma_0=20$ to 
$\Gamma=5$ in a distance $z=3 \times 10^{17}$ cm, following a $\Gamma\propto z^{-2}$ profile. The inlet has a radius of $5\times 10^{16} cm$. The injected EED is a power law  with  
slope $p=2$, 
%and  $\gamma_{max}=5\times 10^6$
the magnetic field at the inlet is $B=0.015 $ G, 
and the jet power is $L_{jet}=2.2 \times 10^{44}$ erg s$^{-1}$.
The solid lines represent the synchrotron, Compton, and total luminosity.
The broken  lines represent the synchrotron and Compton luminosity of the
fastest fifth of the flow, while the dotted lines that of the rest of it.
Lower panel: The same with reduced jet power,    
$L_{jet}=1.6 \times 10^{44}$ erg s$^{-1}$. 
In  both panels the data are the same as in Figure \ref{m87sed}.}
\label{decelm87}
\end{figure}

We apply this  model to the core of M87, assuming a deceleration
 from $\Gamma_0=20$ to $\Gamma=5$ over a  
distance of $3 \times 10^{17}$ cm. We plot the SED as seen at
an angle $\theta=13^{\circ}$
for two different jet luminosities in Figure \ref{decelm87}. In both cases 
most of the observed power comes from the slower part of the flow
 (dotted lines), although the TeV flux is UC emission from the fast base of
the flow (broken lines). 
%The TeV flux is boosted relative to the homogeneous model, 
%because the low energy seed photons required for its production are mostly 
%produced in the slower part of the flow and their energy density is  
%'seen' relativistically boosted in the frame of fast upstream flow 
%in which the TeV electrons are found.   
Note that, as in  homogeneous models, a decrease of the jet 
luminosity (lower panel) results in a steeper - when lower  X-ray SED, 
in agreement with observations. 

\begin{figure}
\epsscale{0.8}
\plotone{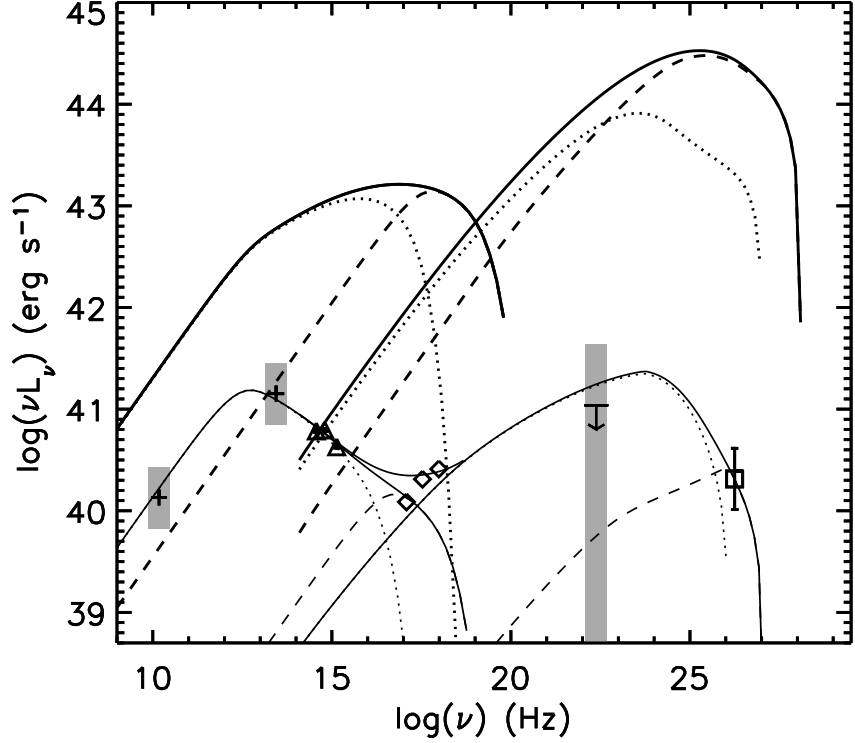}
\caption{The SED of the same decelerating flow as in the upper panel of Figure \ref{decelm87}, as seen under an angle of $\theta=13^{\circ}$ (thin  lines) and
 $\theta=1/\Gamma_0=2.9^{\circ}$  (thick  lines). 
As in Figure \ref{decelm87}, 
the broken  lines represent the synchrotron and Compton luminosity of the
fastest fifth of the flow, while the dotted lines that of the rest of it.
Note how the fast part of the flow dominates the total luminosity of
both the synchrotron and Compton components at  
$\theta=1/\Gamma_0=2.9^{\circ}$.
The data are the same as in Figure \ref{m87sed}.}
\label{tevblazars}
\end{figure}

To check if, in addition to reproducing the 
SED, this model can reproduce the qualitative characteristics 
of  TeV BLs, we plot in Figure \ref{tevblazars} in thick lines the same SED 
viewed at  
$\theta=1/\Gamma_0=2.9^{\circ}$ 
(for comparison we re-plot in thin lines the SED viewed at
  $\theta=13^{\circ}$). While at
  $\theta=13^{\circ}$,  $\nu_s\sim 10^{12.5}$ Hz 
corresponds to the EED cooling break, at  
 $\theta=2.9^{\circ}$ the increased beaming in the fast base of the 
 flow dominates the high energy emission,  resulting in 
   $\nu_s\approx 10^{17}$ Hz. 
%Instrumental to this is the contribution of the faster base of the flow to the synchrotron luminosity, that  increases to be practically equal to that of the rest of the flow.  
The IC flux  increases more than the synchrotron, 
due mostly to UC emission from the fast base of the flow,
resulting in a Compton dominance  $C_d\approx 20$, comparable  to that of   
 1ES 1426+428, the most Compton dominated TeV BL
(Costamante et al. 2003).

\section{Discussion  \label{section:conclusions}}

The  1998/99  TeV flux of M87
is higher than the  2003/04  flux by over a factor 3 
 at  $ \gtrsim  3 \sigma$ significance (Beilicke et al. 2005).
Based on this  and on optical and X-ray variability constraints,
we argued that the source of the  TeV emission of M87  is its core,
as has been previously predicted both in the context of
 leptonic \citep{bai01}  and hadronic models \citep{protheroe03}. 
 We excluded knot A because of its large size and  weak or absent  
variability, and HST 1 because its optical and X-ray flux increased, while
the TeV (HESS) flux remained constant or slightly decreased between 
2003 and 2004.
  TeV emission from either the extended jet of M87 \citep{stawarz04} or 
from hadronic processes in the giant elliptical galaxy itself 
\citep{pfrommer03} also do not  agree with the observed TeV variability.  
\cite{stawarz05} used the fact that the TeV emission varies to set 
an upper limit on the actual TeV emission of knot A (lower that the lowest
 TeV flux observed), and, through this,
a lower limit on its magnetic field.
%, assuming that the TeV flux of knot A
%is due to  inverse-Comptonized starlight from the surrounding galaxy. 

A homogeneous core model is not favored, because  it cannot
reproduce the core SED and is inconsistent with the unified scheme.
A decelerating relativistic flow  (GK03b)
satisfies both the SED modeling and unification requirements.
This is in contrast to both Cen A and NGC 6251, that have be modeled
 as  homogeneous sources \citep{chiaberge01, chiaberge03}. 
Neither of these sources is superluminal, 
suggesting a jet  inclination
larger than that of  M87  and,  in the context of 
a decelerating flow, a  SED  dominated  by
the slower part of the jet. Since this is characterized  by small velocity 
gradients, their SEDs  can be accommodated by  
homogeneous models. 
An orientation difference also explains the lack of TeV emission
from these sources: their TeV emission
is more tightly beamed along the jet axis and points away from our line of 
sight.

The postulated deceleration could be due to  entrainment of external matter
(e.g. Bowman, Leahy, \&  Komissarov 1996) or  Compton drag  
(e.g. Melia \& K\"onigl 1989) on a photon field 
external to the high energy emitting jet, such as  in the fast spine -- slow 
sheath 
model of  Ghisellini, Tavecchio, \&  Chiaberge (2005).   Both 
longitudinal (GK03b) and lateral 
(Ghisellini et al. 2005) velocity gradients can be present 
and quantifying their  importance will require
detailed variability modeling.
We note that arguments  for further deceleration  have been presented 
for FR I 
jets at $\sim 1-10 $ Kpc scales \citep{laing02}, and for  powerful 
FR II  and quasars jets at  $\sim 100$'s of Kpc 
\citep{georganopoulos04}. 
%Understanding the nature  of deceleration as a function
%of jet power will, hopefully, 
%contribute to resolving the reasons for  the FR I - FR II dichotomy.

We end with a cautionary note: if M87 were at a distance similar to 
that of  TeV BLs, the core and knot HST-1 would be seen as a single  
unresolved source in both optical and X-ray energies.
Modeling the broadband SED as being produced at a single
site could then result to unrealistic jet descriptions.
%This danger  is not alleviated even if broadband variability is observed:
%if a  component  varies by  a certain frequency - dependent amount, 
%the variation is  reflected in the total SED, strengthening our false
%impression that we are dealing with a single source. In such a case, 
%depending on the relative contribution of each component, 
% orphan TeV (e.g. Krawczynski et al. 2004) or X-ray flares could be observed.

\acknowledgments
%are these correct?
We thank the referee for his/her suggestions and Dan Harris for insightful discussions.
Our research  is supported by NASA LTSA grants NAG5-9997 and 
NNG 05-GD63DG,  INTEGRAL grant AO2-220080, 
as well as HST grants GO-9142, GO-9847, and GO-9705.

%\appendix

\clearpage

\end{document}